%% file: tradeoff_prl_1.tex
\documentclass[twocolumn,showpacs,preprintnumbers,amsmath,amssymb]{revtex4}

\usepackage{graphicx}
\usepackage{dcolumn}
\usepackage{bm}
\usepackage{theorem}

\input{preferences_ieee}

\input{commands}

\newcommand{\nn}{\nonumber}
\newcommand{\beq}{\begin{eqnarray}}
\newcommand{\eeq}{\end{eqnarray}}

\begin{document}


\title{A Four-Round LOCC Protocol Outperforms All Two-Round Protocols in Reducing the Entanglement Cost for A Distributed Quantum Information Processing}

\author{Eyuri Wakakuwa}
\email{wakakuwa@quest.is.uec.ac.jp}
\affiliation{Graduate School of Information Systems, The University of Electro-Communications, Japan}%

\author{Akihito Soeda}
\author{Mio Murao}%
  \altaffiliation[Also at ]{Institute for Nano Quantum Information Electronics, The University of Tokyo, Japan}
\affiliation{%
Department of Physics, Graduate School of Science, The University of Tokyo, Japan
}%

\date{\today}

\begin{abstract}

We prove that there is a trade-off relation between the entanglement cost and the number of rounds of communication, for two distant parties to accomplish a  bidirectional quantum information task by local operations and classical communication (LOCC). We consider an implementation of a class of two-qubit controlled-unitary gate by LOCC assisted by shared entanglement, in an information theoretical scenario  of asymptotically many input pairs and vanishingly small error. We prove the trade-off relation by showing that one ebit of entanglement per pair is necessary to be consumed for implementing the unitary by any two-round protocol, whereas the entanglement cost by a four-round protocol is strictly smaller than one ebit per pair.
\end{abstract}

\pacs{03.67.Bg, 03.67.Mn}
                       
                                                      
\maketitle


\section{Introduction}

When two distant parties collaborate to perform a distributed quantum information processing, it is necessary to communicate some information with each other. If the communication is restricted to be transmission of classical bits, it may also be necessary to make use of  some entanglement shared in advance, depending on the task. Entanglement and classical communication are thus regarded as resources for distributed quantum information processing, and minimizing the cost of those resources has been one of the central issues in quantum information theory.

A relatively unexplored question about distributed quantum information processing is how the performance of a protocol to accomplish a task depends on the number of rounds of communication in the protocol \cite{eric14}. It has been known that the performance of a protocol with more than one round of communication is strictly better than that of any protocol with only one round of communication, for several tasks such as entanglement distillation \cite{bennett96}, quantum key distribution \cite{daniel03}, state discrimination \cite{scott07,xin08,masaki08} and hypothesis testing \cite{masaki10,masaki14,masaki15}. However, few example of tasks is known for which an $r'$-round protocol outperforms any $r$-round protocol and $2\leq r<r'$, with the exception of the result of \cite{xin08}. Moreover, to our knowledge, it is not known whether there exists a trade-off relation between the entanglement cost and the number of rounds of a protocol for a ``genuinely bidirectional'' task, which cannot be accomplished by any protocol with only one round of communication.

In this paper, we investigate implementation of a bipartite unitary gate by LOCC (local operations and classical communication) assisted by shared entanglement, in an information theoretical scenario introduced in \cite{waka15_dqc_v2}. We prove that, for a class of two-qubit controlled-unitary gates, a four-round protocol outperforms all two-round protocols in reducing the entanglement cost. Thus we provide a first example of genuinely bidirectional tasks for which there is a trade-off relation between the entanglement cost and the number of rounds of communication. It is different from the trade-off relation between the entanglement cost and the {\it classical communication cost}, which is known to exists, e.g., for remote state preparation \cite{bennett05,anura03,bennett01,devetak01}.

This paper is organized as follows. In Section \ref{sec:def}, we introduce definitions of the problem. We present the main result and the proof in Section \ref{sec:respro}. Conclusions are given in Section \ref{sec:conclusion}. Some technical parts of the proof of the main result are presented in Appendices.\\

{\it Notations.} $|\Phi_d\rangle$, $|\Phi_{K_n}\rangle$ and $|\Phi_{L_n}\rangle$ represent the maximally entangled state with the Schmidt rank $d,K_n,L_n\in{\mathbb N}$, respectively. $\pi_d$ is the maximally mixed state of rank $d$. The fidelity and the trace distance between two quantum states $\rho$ and $\sigma$ are  defined as $F(\rho,\sigma):=({\rm Tr}[\sqrt{\sqrt{\rho}\sigma\sqrt{\rho}}])^2$ and $\|\rho-\sigma\|_1:={\rm Tr}[\sqrt{(\rho-\sigma)^2}]$, respectively. We abbreviate $F(\rho,|\psi\rangle\!\langle\psi|)$ as $F(\rho,|\psi\rangle)$. For a quantum operation $\mathcal E$, we abbreviate ${\mathcal E}(|\psi\rangle\!\langle\psi|)$ as ${\mathcal E}(|\psi\rangle)$. $\log{x}$ represents the base $2$ logarithm of $x$.

\section{Definitions}\label{sec:def}

In this section, we describe a task that we analyze in this paper, and present a definition of a trade-off relation between the entanglement cost and the number of rounds.

Suppose Alice and Bob are given a sequence of bipartite quantum states $|\psi_{i_1}\rangle^{AB}\cdots|\psi_{i_n}\rangle^{AB}$, generated by an i.i.d. quantum information source of an ensemble $\{p_i,\psi_i\}_i$. We assume that the source is completely mixed,  i.e.,
\begin{eqnarray}
\sum_ip_i|\psi_i\rangle\!\langle\psi_i|^{AB}=\pi_d^{A}\otimes\pi_d^{B}.\nonumber
\end{eqnarray}
Alice and Bob perform the same bipartite unitary $U^{AB}$ on each of $|\psi_{i_1}\rangle^{AB},\cdots,|\psi_{i_n}\rangle^{AB}$ by LOCC using a resource state $\Phi_{K_n}^{A_0B_0}$, in such a way that the average error vanishes in the limit of $n\rightarrow\infty$.  Following the formulation of the Schumacher compression \cite{schumacher95}, we assume that Alice and Bob do not know the ensemble $\{p_i,\psi_i\}_i$, but know that the average state is completely mixed.  An equivalent task is that Alice and Bob apply $(U^{AB})^{\otimes n}$ on $(|\Phi_d\rangle^{AR_A}|\Phi_d\rangle^{BR_B})^{\otimes n}$ by LOCC using a resource state $\Phi_{K_n}^{A_0B_0}$. Here, $R_A$ and $R_B$ are imaginary reference systems that are inaccessible to Alice and Bob.

In general, a two-party LOCC protocol consists of concatenation of one party performing a local measurement and communicating a classical message to another. The number of concatenation is called the {\it number of rounds} of the protocol. For example, a two-round protocol proceeds as follows: Alice first performs a measurement and communicates the outcome to Bob; Bob then performs a measurement and communicates the outcome to Alice; and, finally, Alice performs an operation.

A rigorous definition of the entanglement cost of a unitary is given below.

\begin{dfn}\label{definition:theprotocol} (Definition 1 in \cite{waka15_dqc_v2})
Let $U$ be a bipartite unitary acting on two $d$-dimensional quantum systems $A$ and $B$. Let Alice and Bob have quantum registers $\{A_0,A_1\}$ and $\{B_0,B_1\}$, respectively, and let ${\mathcal M}_n$ be a quantum operation from $A^nA_0\otimes B^nB_0$ to $A^nA_1\otimes B^nB_1$. ${\mathcal M}_n$ is called an $(r,n,\epsilon)$-protocol for implementing $U$ if  ${\mathcal M}_n$ is an $r$-round LOCC that satisfies
\begin{eqnarray}
F(\rho({\mathcal M_n}), |\Psi_U\rangle^{\otimes n}|\Phi_{L_n}\rangle^{A_1B_1})\geq1-\epsilon,
\label{eq:fidelityn}
\end{eqnarray}
where
\begin{eqnarray}
|\Psi_U\rangle:=U^{AB}|\Phi_d\rangle^{AR_A}|\Phi_d\rangle^{BR_B}\nonumber
\end{eqnarray}
and
\begin{eqnarray}
\rho({\mathcal M}_n):={\mathcal M}_n(|\Phi_{d}^{AR_A}\rangle^{\otimes n}|\Phi_{d}^{BR_B}\rangle^{\otimes n}|\Phi_{K_n}\rangle^{A_0B_0}).\label{eq:deffinst}
\end{eqnarray}
The entanglement cost of ${\mathcal M}_n$ is defined by $\log{K_n}-\log{L_n}$.
\end{dfn}

\begin{dfn}\label{definition:achievablerate}
A rate $E$ is said to be achievable by an $r$-round protocol for implementing $U$ if, for any $\epsilon>0$, there exists $n_\epsilon$ such that for any $n\geq n_\epsilon$, we find an $(r,n,\epsilon)$-protocol for implementing $U$ with the entanglement cost $nE$. For a technical reason, we additionally require that
\begin{align}
\lim_{\epsilon\rightarrow0}\epsilon\cdot n_\epsilon^4=0.\label{eq:solace}
\end{align}
The entanglement cost of $U$ by $r$-round protocols is defined as
\begin{eqnarray}
E_r(U)&:=&{\rm inf}\{E\:|\:E\text{ is achievable by an }r\text{-round}\nonumber\\
&&\;\;\;\;\;\;\;\;\;\text{ protocol for implementing }U\}.\nonumber
\end{eqnarray}
\end{dfn}

The main focus of this paper is whether there is a trade-off relation between the entanglement cost and the number of rounds for implementing a bipartite unitary. In considering ``trade-off relation'', we compare the entanglement cost of a unitary by $r$-round protocols and that by an $r'$-round protocol ($r<r'$). If the latter is strictly smaller than the former, we could say that there exists a trade-off relation between the entanglement cost and the number of rounds. A rigorous definition is as follows: 

\begin{dfn}\label{definition:tradeoff}
{\it There exists a trade-off relation between the entanglement cost and the number of rounds for implementing $U$} if there exists $r,r'\in{\mathbb N}$ such that
\begin{align}
r<r',\;E_r(U)>E_{r'}(U).\nonumber
\end{align}
\end{dfn}

\section{Result and Proof}\label{sec:respro}

We consider a class of two-qubit controlled-phase gate, which  takes the form of
\begin{eqnarray}
U_\theta^{AB}=|0\rangle\!\langle0|^A\otimes I^B+|1\rangle\!\langle1|^A\otimes (e^{i\theta\sigma_z})^B\nonumber
\end{eqnarray}
where
\begin{equation}
\sigma_z=
\begin{pmatrix}
1&0\\
0&-1
\end{pmatrix}
,\; 0<\theta\leq\frac{\pi}{2}.\nonumber
\end{equation}

The main result of this paper is as follows:
\begin{thm}\label{thm:tradeofftheta}
There exists a trade-off relation between the entanglement cost and the number of rounds for implementing $U_\theta$ for any $\theta\in(0,\theta_{\rm max}]$, where $\theta_{\rm max}\in(0,\pi/2]$ is a constant.
\end{thm}
We prove Theorem \ref{thm:tradeofftheta} by showing that the following relations hold for any $\theta\in(0,\theta_{\rm max}]$:
\begin{subequations}
\begin{eqnarray}
&&E_2(U_\theta)\:\geq\:1,\label{subeq:1}\\
&&E_4(U_\theta)<1.\label{subeq:2}
\end{eqnarray}
\end{subequations}
 Inequality (\ref{subeq:1}) is proved in \cite{waka15_dqc_v2} (see  the converse part of Theorem 25 therein),  and an outline of the proof will be presented at the end of this section. We prove Inequality (\ref{subeq:2}) in the following subsections, in which we also derive a stronger relation that
\begin{eqnarray}
\lim_{\theta\rightarrow0}E_4(U_\theta)=0.\label{eq:yamaP}
\end{eqnarray}

\subsection{A Single-Shot Four-Round Protocol}\label{sec:single4}

Let us first describe a single-shot protocol proposed in \cite{ye06} for implementing the following two-qubit unitary gate by four-round LOCC: 
\begin{eqnarray}
{\tilde U}_\theta^{AB}=\cos{\left(\frac{\theta}{2}\right)}\cdot I^A\otimes I^B+i\sin{\left(\frac{\theta}{2}\right)}\cdot \sigma_z^A\otimes \sigma_z^B.\label{eq:atarimae}
\end{eqnarray}
Note that ${\tilde U}_\theta$ is equivalent to $U_\theta$ up to local unitary transformations \cite{kraus01}.

The protocol consists of a concatenation of two two-round protocols. In the first half, Alice and Bob implement ${\tilde U}_\theta$ by using the following state as a shared resource (See Appendix \ref{app:fourround} for the detail):
\begin{eqnarray}
|\phi_\alpha\rangle^{A_0B_0}=\cos{\left(\frac{\alpha}{2}\right)}|0\rangle|0\rangle+i\sin{\left(\frac{\alpha}{2}\right)}|1\rangle|1\rangle.\nonumber
\end{eqnarray}
The protocol is probabilistic and the success probability is given by
\begin{eqnarray}
p(\alpha,\theta)=\frac{\sin^2{\alpha}}{2(1-\cos{\theta}\cos{\alpha})}\nonumber
\end{eqnarray}
If the protocol succeeds, ${\tilde U}_\theta$ is implemented on the input pair as desired, in which case Alice and Bob do nothing in the second half of the protocol. If it fails, then another controlled-unitary  gate ${\tilde U}_{\theta'}$ is applied to the input  state. In that case, Alice and Bob continue to implement ${\tilde U}_{\theta-\theta'}$ by a deterministic protocol proposed in \cite{eisert00} in the second half, which consumes one Bell pair. Note that ${\tilde U}_{\theta-\theta'}{\tilde U}_{\theta'}={\tilde U}_\theta$. Thus the protocol succeeds in implementing ${\tilde U}_\theta$ in total, regardless of the failure in the intermediate step. The average entanglement cost, measured by entanglement entropy, is given by
\begin{eqnarray}
{\bar E}(\alpha,\theta)&=&1-p(\alpha,\theta)+h\left(\cos^2(\alpha/2)\right),\nonumber
\end{eqnarray}
 where $h$ is the binary entropy defined by
\begin{eqnarray}
h(x):=-x\log{x}-(1-x)\log{(1-x)}.\nonumber
\end{eqnarray}

Define
\begin{align}
\alpha_\theta:=\sqrt{\theta},\:p_\theta:=p(\alpha_\theta,\theta),\:E_\theta:={\bar E}(\alpha_\theta,\theta).\nonumber
\end{align}
It is straightforward to verify that $E_\theta$ is a continuous function of $\theta\in(0,\pi/2]$. As we prove in Appendix \ref{app:prftheta0}, the function satisfies
\begin{eqnarray}
\lim_{\theta\rightarrow0}E_\theta=0.\label{eq:theta0}
\end{eqnarray}
Thus there exists a constant $\theta_{\rm max}\in(0,\pi/2]$ such that we have
\begin{align}
E_\theta<1\label{eq:theta1}
\end{align}
for all $\theta\in(0,\theta_{\rm max}]$.

\subsection{An $n$-Shot Protocol}

Let us consider an $n$-shot protocol for implementing ${\tilde U}_\theta$. Fix arbitrary $\delta>0$ and $n\in{\mathbb N}$. The protocol proceeds as follows:

\begin{itemize}
\item[I-1.] Alice and Bob initially share $n$ copies of $|\phi_{\alpha_\theta}\rangle$ and $n(1-p_\theta+\delta)$ Bell pairs.
\item[I-2.] By using $n$ copies of $|\phi_\alpha\rangle$ as resources, they perform ${\tilde U}_\theta$ on each  of the input sequence by the first half of the protocol described in Section \ref{sec:single4}. Either of the following two events will occur:
\begin{enumerate}\renewcommand{\labelenumi}{(\alph{enumi})}
\item The number of pairs for which ${\tilde U}_\theta$ has been applied is not smaller than $n(p_\theta-\delta)$. ${\tilde U}_{\theta'}$ has been applied on the other pairs, the number of which is not greater than $n(1-p_\theta+\delta)$.  
\item The number of pairs for which ${\tilde U}_\theta$ has been applied is smaller than $n(p_\theta-\delta)$.
\end{enumerate}
Continue to the next step if (a) has occurred.
\item[I-3.] By using $n(1-p_\theta+\delta)$ Bell pairs, they perform ${\tilde U}_{\theta-\theta'}$ by the second half of the protocol described in Section \ref{sec:single4}, on pairs for which ${\tilde U}_{\theta'}$ has been applied.
\end{itemize}

Let ${\mathcal M}_n'$ be a quantum operation that represents Step I-2 and I-3, and suppose the input state is
\begin{align}
|\Psi_n\rangle^{A^nB^n}:=|\psi_{1}\rangle^{AB}\cdots|\psi_{n}\rangle^{AB}.\nonumber
\end{align}
The total error is evaluated as follows. Let $\epsilon_n$ be the probability that (b) occurs in Step I-2, and let $\tau_{(b)}$ be the state obtained when (b) occurs. If (a) occurs in Step I-2, the final state is exactly equal to the target state $|\Psi_{n,{\rm tar}}\rangle:={\tilde U}^{\otimes n}|\Psi_n\rangle$. Thus the final state is, in total, given by
\begin{eqnarray}
&&{\mathcal M}_n'\left(|\Psi_n\rangle|\phi_{\alpha_\theta}\rangle^{\otimes n}|\Phi_2\rangle^{\otimes n(1-p_\theta+\delta)}\right)\nonumber\\
&&=(1-\epsilon_n)|\Psi_{n,{\rm tar}}\rangle\!\langle\Psi_{n,{\rm tar}}|+\epsilon_n\tau_{(b)},\nonumber
\end{eqnarray}
which leads to
\begin{eqnarray}
&&\left\|{\mathcal M}_n'\left(|\Psi_n\rangle|\phi_{\alpha_\theta}\rangle^{\otimes n}|\Phi_2\rangle^{\otimes n(1-p_\theta+\delta)}\right)\right.\nonumber\\
&&\;\;\;\;\;\;\left.-{\tilde U}_{\theta}^{\otimes n}|\Psi_n\rangle\!\langle\Psi_n|{\tilde U}_{\theta}^{\dagger \otimes n}\right\|_1\nonumber\\
&=&\epsilon_n\left\||\Psi_{n,{\rm tar}}\rangle\!\langle\Psi_{n,{\rm tar}}|-\tau_{(b)}\right\|_1\leq2\epsilon_n.\label{eq:ashla}
\end{eqnarray}

The law of large numbers implies $\lim_{n\rightarrow\infty}\epsilon_n=0$. It is proved in \cite{ahlswede80} that there exists an $n$-independent positive constant $c_\theta$ such that
\begin{align}
\epsilon_n\leq \exp{(-c_\theta\delta^2n)}\label{eq:agentcho}
\end{align}
for any $\delta$ and $n$.

\subsection{Proof of Inequality (\ref{subeq:2})}

We prove that
\begin{align}
E_4(U_\theta)\leq E_\theta\;\;\text{for any }\theta\in(0,\pi/2].\nonumber
\end{align}
This yields Inequality (\ref{subeq:2}) for $\theta\in(0,\theta_{\rm max}]$ due to (\ref{eq:theta1}), as well as (\ref{eq:yamaP}) due to (\ref{eq:theta0}). Note that the local unitary equivalence of $U_\theta$ and ${\tilde U}_\theta$ implies $E_4(U_\theta)=E_4({\tilde U}_\theta)$. Thus we prove in the following that $E_4({\tilde U}_\theta)\leq E_\theta$ for any $\theta\in(0,\pi/2]$. We denote $h(\cos^2(\alpha_\theta/2))$ simply by $h_\theta$.

Fix arbitrary $\delta>0$ and $n\in{\mathbb N}$, and consider the following protocol for implementing ${\tilde U}_\theta$ with the entanglement cost $n(E_\theta+2\delta)$.
\begin{itemize}
\item[I\!I-1.] Alice and Bob initially share a maximally entangled state with Schmidt rank $K_n=2^{n(E_\theta+2\delta)}$.
\item[I\!I-2.] Alice and Bob transforms the resource entanglement to $n(E_\theta+2\delta)$ copies of Bell pairs by local unitary operations.
\item[I\!I-3.] By entanglement dilution \cite{bennett966}, they transform $n(h_\theta+\delta)$ copies of Bell pairs to a state $\omega_n$ which is close to $|\phi_{\alpha_\theta}\rangle^{\otimes n}$.
\item[I\!I-4.] Alice and Bob perform ${\mathcal M}_n'$ by using $\omega_n$ and the remaining $n(1-p_\theta+\delta)$ Bell pairs  as resource.
\end{itemize}

Let ${\mathcal M}_n$ be a quantum operation that represents Step  I\!I-2$\sim$4, and define 
\begin{align}
\epsilon_n':=\left\||\omega_n\rangle\!\langle\omega_n|-|\phi_{\alpha_\theta}\rangle\!\langle\phi_{\alpha_\theta}|^{\otimes n}\right\|_1.\label{eq:ishedead}
\end{align}
By definition, we have
\begin{eqnarray}
{\mathcal M}_n\left(|\Psi_n\rangle|\Phi_{K_n}\rangle\right)={\mathcal M}_n'\left(|\Psi_n\rangle|\omega_n\rangle|\Phi_2\rangle^{\otimes n(1-p_\theta+\delta)}\right).\nonumber
\end{eqnarray}
A simple calculation then yields 
\begin{align}
\left\|{\mathcal M}_n\left(|\Psi_n\rangle|\Phi_{K_n}\rangle\right)-{{\tilde U}_\theta}^{\otimes n}|\Psi_n\rangle\!\langle\Psi_n|{{\tilde U}_\theta}^{\dagger \otimes n}\right\|_1\leq2\epsilon_n+\epsilon_n'\label{eq:koisurubonjin}
\end{align}
from (\ref{eq:ashla}) (see Appendix \ref{app:orionshobo}).

Since this relation holds for any $|\Psi_n\rangle\in({\mathcal H}^{A}\otimes{\mathcal H}^{B})^{\otimes n}$, it follows that
\begin{align}
\left\|{\mathcal M}_n\left(|\Phi_2^{AR_A}\rangle^{\otimes n}|\Phi_2^{BR_B}\rangle^{\otimes n}|\Phi_{K_n}\rangle\right)-|\Psi_{{\tilde U}_\theta}\rangle\!\langle\Psi_{{\tilde U}_\theta}|^{\otimes n}\right\|_1\nonumber\\
\leq2\epsilon_n+\epsilon_n'.\nonumber
\end{align}
As we prove in Appendix \ref{app:prfdisti}, there exists an $n$-independent positive constant $c'_\theta$ such that
\begin{align}
\epsilon_n'\leq 2\exp{\left(-\frac{c_\theta'\delta^2n}{2}\right)}\label{eq:epsilonpr}
\end{align}
for any $\delta>0$ and  $n\in{\mathbb N}$. This ensures Condition (\ref{eq:solace}) combined with (\ref{eq:agentcho}), noting that the fidelity and the trace distance are related as $F(\rho,\sigma)\geq1-\|\rho-\sigma\|_1$ (see e.g. \cite{wildetext}). Since $\delta>0$ can be arbitrarily small, we obtain $E_4({\tilde U}_\theta)\leq E_\theta$.
\hfill$\blacksquare$

\subsection{Outline of the Proof of Inequality (\ref{subeq:1})}
\label{sec:outprfe21}

Let us first consider an arbitrary bipartite unitary $U$ acting on two $d$-level systems $A$ and $B$. Define a ``tripartite'' state
\begin{align}
|\Psi_U\rangle^{AR_A(\!BR_B\!)}:=(U^{AB}\otimes I^{R_AR_B})|\Phi_d\rangle^{AR_A}|\Phi_d\rangle^{BR_B}\nn
\end{align}
by regarding $B$ and $R_B$ as a single system. Consider a task in which $n$ copies of $|\Psi_U\rangle^{AR_A(\!BR_B\!)}$ is transformed by a random unitary operation on $A^n$ to a Markov state conditioned by $B^n$,  i.e., a state that satisfies $I(A^n:B^nR_B^n|R_A^n)=0$ \cite{hayden04}. In particular, suppose $2^{nR}$ unitary operations are randomly applied on $A^n$ with the uniform distribution, and the trace distance between the final state and a Markov state vanishes in the limit of $n\rightarrow\infty$. The infimum ratio $R$ satisfying this condition is called the {\it Markovianizing cost of $U$}, and is denoted by $M(U)$ \cite{waka15_markov,waka15_dqc_v2}. The following proposition states that $M(U^\dagger)$ is a lower bound on the entanglement cost for implementing a bipartite unitary by a two-round protocol.

\begin{prp}\label{thm:misoptach}(Corollary of the converse part of Theorem 25 in \cite{waka15_dqc_v2})
A rate $E$ is achievable by a two-round protocol for implementing $U$ only if $E\geq M(U^\dagger)$, if we require Condition (\ref{eq:solace}) in Definition \ref{definition:achievablerate}.
\end{prp}

The Markovianizing cost of a bipartite unitary is computed as follows. The {\it Petz recovery map} ${\mathcal R}_U:A\rightarrow A(BR_B)$ corresponding to $|\Psi_U\rangle^{AR_A(\!BR_B\!)}$ is defined by
\begin{align}
{\mathcal R}_U(\tau)=(\Psi_U^{A(BR_B)})^{\frac{1}{2}}(\Psi_U^{A})^{-\frac{1}{2}}\tau(\Psi_U^{A})^{-\frac{1}{2}}(\Psi_U^{A(BR_B)})^{\frac{1}{2}}\nn\\
=U^{AB}({\rm Tr}_B[U^{\dagger AB}(\tau^A\otimes I^B)U^{AB}]\otimes\Phi_d^{BR_B})U^{\dagger AB}\nn
\end{align}
for $\tau\in{\mathcal S}({\mathcal H}^A)$ \cite{hayden04}. Define CPTP maps ${\mathcal E}_U$ and ${\mathcal E}_{U,\infty}$ on $A$ by 
\begin{align}
{\mathcal E}_U:={\rm Tr}_{BR_B}\circ{\mathcal R}_U,\;\;\;\;{\mathcal E}_{U,\infty}:=\lim_{N\rightarrow\infty}\frac{1}{N}\sum_{n=1}^N{\mathcal E}_U^n,\nn
\end{align}
and consider the state
\begin{eqnarray}
\Phi_{U,\infty}^{AR_A}:={\mathcal E}_{U,\infty}^A(|\Phi_d\rangle\!\langle\Phi_d|^{AR_A}).\nn
\end{eqnarray}
It is proved in \cite{waka15_dqc_v2} that the Markovianizing cost of $U$ is equal to the von Neumann entropy of $\Phi_{U,\infty}^{AR_A}$, i.e.,
\begin{align}
M(U)=S(\Phi_{U,\infty}^{AR_A}).\nn
\end{align} 

For ${\tilde U}_\theta$ defined by (\ref{eq:atarimae}), we have
\begin{eqnarray}
{\mathcal E}_{{\tilde U}_\theta^\dagger}(\tau)=\frac{1+\cos^2{\theta}}{2}\cdot\tau+\frac{1}{2}\sin^2{\theta}\cdot \sigma_z\tau \sigma_z,\nn
\end{eqnarray}
which leads to
\begin{eqnarray}
{\mathcal E}_{{\tilde U}_\theta^\dagger,\infty}(\tau)=\frac{1}{2}(\tau+\sigma_z\tau\sigma_z)=\proj{0}\tau\proj{0}+\proj{1}\tau\proj{1}.\nn
\end{eqnarray}
Hence we have
\begin{align}
\Phi_{{\tilde U}_\theta^\dagger,\infty}^{AR_A}=\frac{1}{2}(|0\rangle\!\langle0|\otimes|0\rangle\!\langle0|+|1\rangle\!\langle1|\otimes|1\rangle\!\langle1|),\nn
\end{align}
which implies $M({\tilde U}_\theta^\dagger)=1$. Therefore, due to Proposition \ref{thm:misoptach}, we obtain Inequality (\ref{subeq:1}).

\section{Conclusion} \label{sec:conclusion}

We considered implementation of a class of two-qubit controlled-unitary gate by local operations and classical communication (LOCC), assisted by shared entanglement. We proved that a four-round protocol outperforms all two-round LOCC protocols in reducing the entanglement cost. Our result provides a first example of  genuinely bidirectional distributed quantum tasks, for which there exists a trade-off relation between the entanglement cost and the number of rounds of communication.

\begin{acknowledgments}
This work is supported by the Project for Developing Innovation Systems of MEXT, Japan and JSPS KAKENHI (Grant No.~23540463, No.~23240001, No.~26330006, and No.~15H01677).  We also gratefully acknowledge to the ELC project (Grant-in-Aid for Scientific Research on Innovative Areas MEXT KAKENHI (Grant No.~24106009)) for encouraging the research presented in this paper. 
\end{acknowledgments}

\appendix

\section{A probailistic protocol for two-qubit controlled-unitaries}\label{app:fourround}

In this Appendix, we describe a protocol for implementing ${\tilde U}_\theta$ by using resource state
\begin{eqnarray}
|\phi_\alpha\rangle^{A_0B_0}=\cos{\left(\frac{\alpha}{2}\right)}|0\rangle|0\rangle+i\sin{\left(\frac{\alpha}{2}\right)}|1\rangle|1\rangle,\nonumber
\end{eqnarray}
which is proposed in \cite{ye06}. Suppose the input state is $|\psi\rangle^{AB}$. The protocol proceeds as follows:
\begin{enumerate}
\item Alice performs the controlled-$z$ gate
\begin{eqnarray}
U^{A_0A}=\proj{0}^{A_0}\otimes I^A+\proj{1}^{A_0}\otimes \sigma_z^A,\nonumber
\end{eqnarray}
after which the whole state is
\begin{eqnarray}
|\psi_{tot}'\rangle^{A_0B_0AB}&=&\cos{\left(\frac{\alpha}{2}\right)}|0\rangle^{A_0}|0\rangle^{B_0}|\psi\rangle^{AB}\nonumber\\
&&\!\!\!\!\!\!\!\!+i\sin{\left(\frac{\alpha}{2}\right)}|1\rangle^{A_0}|1\rangle^{B_0}\sigma_z^A|\psi\rangle^{AB}.\nonumber
\end{eqnarray}
\item Alice performs a projective measurement on $A_0$ with basis $\{|+\rangle,|-\rangle\}$, and sends the outcome to Bob.
\item Bob performs $I$ or $\sigma_z$ on $B_0$ depending on the measurement outcome. The whole state is then
\begin{eqnarray}
|\psi_{tot}''\rangle^{B_0AB}&=&\cos{\left(\frac{\alpha}{2}\right)}|0\rangle^{B_0}|\psi\rangle^{AB}\nonumber\\
&&+i\sin{\left(\frac{\alpha}{2}\right)}|1\rangle^{B_0}\sigma_z^A|\psi\rangle^{AB}.\nonumber
\end{eqnarray}
\item Alice performs the controlled-$z$ gate
\begin{eqnarray}
U^{B_0B}=\proj{0}^{B_0}\otimes I^B+\proj{1}^{B_0}\otimes \sigma_z^B,\nonumber
\end{eqnarray}
after which the whole state is
\begin{eqnarray}
|\psi_{tot}'''\rangle^{B_0AB}&=&\cos{\left(\frac{\alpha}{2}\right)}|0\rangle^{B_0}|\psi\rangle^{AB}\nonumber\\
&&\!\!\!\!\!\!\!\!\!+i\sin{\left(\frac{\alpha}{2}\right)}|1\rangle^{B_0}(\sigma_z^A\otimes \sigma_z^B)|\psi\rangle^{AB}.\nonumber
\end{eqnarray}
\item Bob performs a projective measurement on $B_0$ with basis $\{|\chi\rangle/\langle\chi|\chi\rangle^{1/2},|\chi^\perp\rangle/\langle\chi^\perp|\chi^\perp\rangle^{1/2}\}$, and sends the outcome to Alice. Here, $|\chi\rangle$ and $|\chi^\perp\rangle$ are supernormalized state vectors defined by
\begin{eqnarray}
&&|\chi\rangle:=\frac{\cos{(\theta/2)}}{\cos{(\alpha/2)}}|0\rangle+\frac{\sin{(\theta/2)}}{\sin{(\alpha/2)}}|1\rangle,\nonumber\\
&&|\chi^\perp\rangle:=\frac{\sin{(\theta/2)}}{\sin{(\alpha/2)}}|0\rangle-\frac{\cos{(\theta/2)}}{\cos{(\alpha/2)}}|1\rangle.\nonumber
\end{eqnarray}
\end{enumerate}
If the measurement outcome corresponding to $|\chi\rangle$ is obtained, the state becomes
\begin{eqnarray}
&&\!\!\!\!\!\!|\psi^s\rangle^{AB}=\langle\chi|\psi_{tot}'''\rangle\nonumber\\
&&=\cos{\left(\frac{\theta}{2}\right)}|\psi\rangle^{AB}+i\sin{\left(\frac{\theta}{2}\right)}(\sigma_z^A\otimes \sigma_z^B)|\psi\rangle^{AB}\nonumber
\end{eqnarray}
as desired. The success probability is given by 
\begin{eqnarray}
p(\alpha,\theta)=\frac{|\langle\chi|\psi_{tot}'''\rangle|^2}{\langle\chi|\chi\rangle}=\frac{1}{\langle\chi|\chi\rangle}=\frac{\sin^2{\alpha}}{2(1-\cos{\theta}\cos{\alpha})}.\nonumber
\end{eqnarray}
If the complementary outcome is obtained, then the state changes
\begin{eqnarray}
&&\!\!\!\!\!\!\!\!\!|\psi^f\rangle^{AB}=\langle\chi^\perp|\psi_{tot}'''\rangle\nonumber\\
&&=\frac{\sin{(\theta/2)}}{\tan{(\alpha/2)}}|\psi\rangle^{AB}+i\frac{\cos{(\theta/2)}}{\tan{(\alpha/2)}^{-1}}(\sigma_z^A\otimes \sigma_z^B)|\psi\rangle^{AB},\nonumber
\end{eqnarray}
up to normalization condition. It is straightforward to verify that the normalized state satisfies
\begin{align}
\frac{|\psi^f\rangle^{AB}}{\||\psi^f\rangle^{AB}\|}={\tilde U}_{\theta'}|\psi\rangle^{AB}\nonumber
\end{align}
with $\theta'$ defined by
\begin{eqnarray}
\tan{\left(\frac{\theta'}{2}\right)}=\frac{\tan^2{(\alpha/2)}}{\tan{(\theta/2)}}.\nonumber
\end{eqnarray}

\section{Proof of Equality (\ref{eq:theta0}) and Inequality (\ref{eq:koisurubonjin})}

\subsection{Equality (\ref{eq:theta0})}\label{app:prftheta0}

By definition, we have
\begin{eqnarray}
&&E_\theta=1-p_\theta+h(\cos^2(\sqrt{\theta}/2)),\label{eq:ethetafor}\\
&&p_\theta=\frac{\sin^2{\sqrt{\theta}}}{2(1-\cos{\theta}\cos{\sqrt{\theta}})}.\nonumber
\end{eqnarray}
It is straightforward to verify that
\begin{align}
\lim_{\theta\rightarrow0}h(\cos^2(\sqrt{\theta}/2))=0.\label{eq:hoshizora}
\end{align}
For $\theta\approx0$, we have
\begin{eqnarray}
&&\sin^2{\sqrt{\theta}}=\theta+O(\theta^2),\nonumber\\
&&\cos{\theta}=1-\frac{1}{2}\theta^2+O(\theta^4),\nonumber\\
&&\cos{\sqrt{\theta}}=1-\frac{1}{2}\theta+O(\theta^2),\nonumber\\
&&\cos{\theta}\cos{\sqrt{\theta}}=1-\frac{1}{2}\theta+O(\theta^2).\nonumber
\end{eqnarray}
Thus we have
\begin{align}
p_\theta=\frac{\theta+O(\theta^2)}{2\left(\frac{1}{2}\theta+O(\theta^2)\right)}=1+O(\theta),\nonumber
\end{align}
which leads to
\begin{align}
\lim_{\theta\rightarrow0}p_\theta=1.\label{eq:pthetaconv}
\end{align}
From (\ref{eq:ethetafor}), (\ref{eq:hoshizora}) and (\ref{eq:pthetaconv}), we obtain (\ref{eq:theta0}).\hfill$\blacksquare$

\subsection{Inequality (\ref{eq:koisurubonjin})}\label{app:orionshobo}

We obtain Inequality (\ref{eq:koisurubonjin}) as
\begin{eqnarray}
&&\left\|{\mathcal M}_n(|\Psi_n\rangle|\Phi_{K_n}\rangle)-{\tilde U}_\theta^{\otimes n}|\Psi_n\rangle\!\langle\Psi_n|{\tilde U}_\theta^{\dagger \otimes n}\right\|_1\nonumber\\
&=&\left\|{\mathcal M}_n'\left(|\Psi_n\rangle|\omega_n\rangle|\Phi_2\rangle^{\otimes n(1-p_\theta+\delta)}\right)\right.\nonumber\\
&&\;\;\left.-{\tilde U}_\theta^{\otimes n}|\Psi_n\rangle\!\langle\Psi_n|{\tilde U}_\theta^{\dagger \otimes n}\right\|_1\nonumber\\
&\leq&\left\|{\mathcal M}_n'\left(|\Psi_n\rangle|\omega_n\rangle|\Phi_2\rangle^{\otimes n(1-p_\theta+\delta)}\right)\right.\nonumber\\
&&\;\;\left.-{\mathcal M}_n'\left(|\Psi_n\rangle|\phi_{\alpha_\theta}\rangle^{\otimes n}|\Phi_2\rangle^{\otimes n(1-p_\theta+\delta)}\right)\right\|_1\nonumber\\
&&+\left\|{\mathcal M}_n'\left(|\Psi_n\rangle|\phi_{\alpha_\theta}\rangle^{\otimes n}|\Phi_2\rangle^{\otimes n(1-p_\theta+\delta)}\right)\right.\nonumber\\
&&\;\;\;\;\;\;\left.-{\tilde U}_\theta^{\otimes n}|\Psi_n\rangle\!\langle\Psi_n|{\tilde U}_\theta^{\dagger \otimes n}\right\|_1\nonumber\\
&\leq&\left\||\Psi_n\rangle\!\langle\Psi_n|\otimes|\omega_n\rangle\!\langle\omega_n|\otimes|\Phi_2\rangle\!\langle\Phi_2|^{\otimes n(1-p_\theta+\delta)}\right.\nonumber\\
&&\;\;\left.-|\Psi_n\rangle\!\langle\Psi_n|\otimes|\phi_{\alpha_\theta}\rangle\!\langle\phi_{\alpha_\theta}|^{\otimes n}\otimes|\Phi_2\rangle\!\langle\Phi_2|^{\otimes n(1-p_\theta+\delta)}\right\|_1\nonumber\\
&&+2\epsilon_n\nonumber\\
&=&\left\||\omega_n\rangle\!\langle\omega_n|-|\phi_{\alpha_\theta}\rangle\!\langle\phi_{\alpha_\theta}|^{\otimes n}\right\|_1+2\epsilon_n\nonumber\\
&=&\epsilon_n'+2\epsilon_n.\nonumber
\end{eqnarray}
Here, the first line follows from the definition of ${\mathcal M}_n'$; the second line due to the triangle inequality for the trace distance; the third line from the monotonicity of the trace distance and Inequality (\ref{eq:ashla}); the forth line because we have $\|\rho\otimes\tau-\sigma\otimes\tau\|_1=\|\rho-\sigma\|_1$; and the fifth line from Inequality (\ref{eq:ishedead}).\hfill$\blacksquare$

\section{Proof of Inequality (\ref{eq:epsilonpr})}\label{app:prfdisti}

\subsection{Typical Subspace}

Define
\begin{align}
\lambda_0=\cos^2\left(\frac{\alpha_\theta}{2}\right),\;\lambda_1=\sin^2\left(\frac{\alpha_\theta}{2}\right),\nonumber
\end{align}
and fix arbitrary $\delta>0$, $n\in{\mathbb N}$. A sequence ${\bm x}=(x_1,\cdots,x_n)\in\{0,1\}^n$ is said to be {\it$\delta$-weakly typical with respect to} $\{\lambda_x\}_{x\in\{0,1\}}$ if it satisfies
\begin{eqnarray}
2^{-n(H(\{\lambda_x\})+\delta)}\leq\prod_{i=1}^n\lambda_{x_i}\leq2^{-n(H(\{\lambda_x\})-\delta)}.\label{eq:kinoshitade}
\end{eqnarray}
Here, $H(\{\lambda_x\})$ is the Shannon entropy of a probability distribution $\{\lambda_x\}_{x\in\{0,1\}}$ defined by
\begin{align}
H(\{\lambda_x\}):=-\sum_{x=\{0,1\}}\lambda_x\log{\lambda_x},\nonumber
\end{align}
and is equal to $h_\theta$. The set of all $\delta$-weakly typical sequences is called the {\it$\delta$-weakly typical set}, and is denoted by ${\mathcal T}_{n,\delta}$. The {\it$\delta$-weakly typical subspace of $({\mathcal H}^{A_0})^{\otimes n}$ with respect to $\phi_{\alpha_\theta}^{A_0}={\rm Tr}_{B_0}[|\phi_{\alpha_\theta}\rangle\!\langle\phi_{\alpha_\theta}|^{A_0B_0}]$} is defined as
\begin{eqnarray}
&&\!\!\!\!\!\!\!\!\!\!\!{\mathcal H}_{n,\delta}:=\nonumber\\
&&\!\!\!\!\!\!{\rm span}\left\{\left.\ket{x_1}\cdots\ket{x_n}\in({\mathcal H}^{A_0})^{\otimes n}\right|(x_1,\cdots,x_n)\in{\mathcal T}_{n,\delta}\right\}.\!\!\!\!\!\!\!\!\!\nonumber
\end{eqnarray}
Let $\Pi_{n,\delta}$ be the projection onto ${\mathcal H}_{n,\delta}\subseteq({\mathcal H}^{A_0})^{\otimes n}$, and let us introduce a notation 
\begin{align}
\lambda_{\bm x}:=\lambda_{x_1}\cdots\lambda_{x_n}.\nonumber
\end{align}
 Abbreviating $(\Pi_{n,\delta}\otimes I^{B_0^n})|\phi_{\alpha_\theta}\rangle^{\otimes n}$ as $\Pi_{n,\delta}|\phi_{\alpha_\theta}\rangle^{\otimes n}$, we have
\begin{align}
{\rm Tr}[\Pi_{n,\delta}(|\phi_{\alpha_\theta}\rangle\!\langle\phi_{\alpha_\theta}|^{\otimes n})]=\sum_{{\bm x}\in{\mathcal T}_{n,\delta}}\lambda_{\bm x}.\label{eq:a}
\end{align}

It is proved in \cite{ahlswede80} that there exists a constant $c>0$, which depends on ${\{\lambda_x\}_x}$, such that for any $\delta>0$ and $n$, we have
\begin{eqnarray}
\sum_{{\bm x}\in{\mathcal T}_{n,\delta}}\lambda_{\bm x}\geq1-\exp{(-c\delta^2n)}.\nonumber
\end{eqnarray}
Denoting this constant by $c_\theta'$, we obtain
\begin{align}
{\rm Tr}[\Pi_{n,\delta}(|\phi_{\alpha_\theta}\rangle\!\langle\phi_{\alpha_\theta}|^{\otimes n})]\geq1-\exp{(-c_\theta'\delta^2n)}\nonumber
\end{align}
from (\ref{eq:a}).

\subsection{Proof of Inequality (\ref{eq:epsilonpr})}
Fix arbitrary $\delta>0$, $n\in{\mathbb N}$, and consider the normalized state $|\omega_n\rangle$ defined by
\begin{eqnarray}
|\omega_n\rangle:=\frac{\Pi_{n,\delta}(|\phi_{\alpha_\theta}\rangle^{\otimes n})}{\sqrt{{\rm Tr}[\Pi_{n,\delta}(|\phi_{\alpha_\theta}\rangle\!\langle\phi_{\alpha_\theta}|^{\otimes n})]}}.
\end{eqnarray}
Due to the gentle measurement lemma (see e.g. Lemma 9.4.1 in \cite{wildetext}), the state satisfies
\begin{eqnarray}
\left\||\omega_n\rangle\!\langle\omega_n|-|\phi_{\alpha_\theta}\rangle\!\langle\phi_{\alpha_\theta}|^{\otimes n}\right\|_1\leq2\exp{\left(-\frac{c_\theta'\delta^2n}{2}\right)}.\nonumber
\end{eqnarray}
By definition, the Schmidt decomposition of $|\omega_n\rangle$ is given by
\begin{eqnarray}
|\omega_n\rangle=\sum_{{\bm x}\in{\mathcal T}_{n,\delta}}\sqrt{\lambda_{\bm x}'}|{\bm x}\rangle|{\bm x}\rangle,\nonumber
\end{eqnarray}
where
\begin{eqnarray}
\lambda_{\bm x}':=\frac{\lambda_{\bm x}}{{\rm Tr}[\Pi_{n,\delta}(|\phi_{\alpha_\theta}\rangle\!\langle\phi_{\alpha_\theta}|^{\otimes n})]}.\nonumber
\end{eqnarray}
From (\ref{eq:kinoshitade}), it follows that
\begin{eqnarray}
\lambda_{\bm x}'\geq2^{-n(H(\{\lambda_x\})+\delta)}.\nonumber
\end{eqnarray}
Thus a uniform distribution on a set $\{1,\cdots,2^{n(H(\{\lambda_x\})+\delta)}\}$ is majorized by a probability distribution $\{\lambda_{\bm x}'\}_{{\bm x}\in{\mathcal T}_{n,\delta}}$. Consequently, due to \cite{nielsen99}, there exists a LOCC protocol that transforms $n(H(\{\lambda_x\})+\delta)$ copies of Bell pairs to $|\omega_n\rangle$ deterministically and exactly.\\\hfill$\blacksquare$

\bibliography{markov.bib}

\end{document}

%% file: preferences_ieee.tex
{\theorembodyfont{\normalfont}
\theoremheaderfont{\normalfont\it}
\newtheorem{thm}{ Theorem}
\newtheorem{dfn}[thm]{ Definition}

\newtheorem{prp}[thm]{ Proposition}}

{\theorembodyfont{\normalfont}
\theoremheaderfont{\normalfont\it}
}

{\theorembodyfont{\normalfont}
\theoremheaderfont{\normalfont\it}
}

{\theorembodyfont{\normalfont}
\theoremheaderfont{\normalfont\it}
}

{\theorembodyfont{\normalfont}
\theoremheaderfont{\normalfont\it}
}

{\theorembodyfont{\normalfont}
\theoremheaderfont{\normalfont\it}
}

%% file: commands.tex
\newcommand{\bra}[1]{\mbox{$\left\langle#1\right|$}}
\newcommand{\ket}[1]{\mbox{$\left|#1\right\rangle$}}

\newcommand{\proj}[1]{\mbox{$\ket{#1}\!\bra{#1}$}}